\documentclass[english,aps,prl,twocolumn,superscriptaddress,floatfix,10pt]{revtex4}
\usepackage[T1]{fontenc}
\usepackage[latin9]{inputenc}
\usepackage{amsthm}
\usepackage{amsmath}
\usepackage{graphicx}

\makeatletter
\@ifundefined{textcolor}{}
{%
 \definecolor{BLACK}{gray}{0}
 \definecolor{WHITE}{gray}{1}
 \definecolor{RED}{rgb}{1,0,0}
 \definecolor{GREEN}{rgb}{0,1,0}
 \definecolor{BLUE}{rgb}{0,0,1}
 \definecolor{CYAN}{cmyk}{1,0,0,0}
 \definecolor{MAGENTA}{cmyk}{0,1,0,0}
 \definecolor{YELLOW}{cmyk}{0,0,1,0}
 }
\numberwithin{equation}{section}
\numberwithin{figure}{section}

\usepackage{chngcntr}
\counterwithout{figure}{section}
\counterwithout{equation}{section}

\makeatother

\usepackage{babel}
\begin{document}

\title{Electrolyte gate-controlled Kondo effect in SrTiO$_{3}$}

\author{Menyoung Lee}

\affiliation{Department of Physics, Stanford University, Stanford, California
94305, USA}

\author{J. R. Williams}

\affiliation{Department of Physics, Stanford University, Stanford, California
94305, USA}

\author{Sipei Zhang}

\affiliation{Department of Chemical Engineering and Materials Science, University
of Minnesota, Minneapolis, Minnesota 55455, USA}

\author{C. Daniel Frisbie}

\affiliation{Department of Chemical Engineering and Materials Science, University
of Minnesota, Minneapolis, Minnesota 55455, USA}

\author{D. Goldhaber-Gordon}

\email{goldhaber-gordon@stanford.edu}

\affiliation{Department of Physics, Stanford University, Stanford, California
94305, USA}

\date{July 30, 2011}
\begin{abstract}
We report low-temperature, high-field magnetotransport measurements
of SrTiO$_{3}$ gated by an ionic gel electrolyte. A saturating resistance
upturn and negative magnetoresistance that signal the emergence of
the Kondo effect appear for higher applied gate voltages. This observation,
enabled by the wide tunability of the ionic gel-applied electric field,
promotes the interpretation of the electric field-effect induced 2D
electron system in SrTiO$_{3}$ as an admixture of magnetic Ti$^{3+}$
ions, i.e. localized and unpaired electrons, and delocalized electrons
that partially fill the Ti $3d$ conduction band.
\end{abstract}


\maketitle

The Coulomb interaction amongst electrons and ions in a solid can
spontaneously generate internal magnetic fields and effective magnetic
interactions. Unexpected magnetic phenomena may emerge whenever we
consider a new system where interactions are important. In recent
years, predictions for and observations of magnetism originating in
the two-dimensional (2D) system of electrons at the interface between
SrTiO$_{3}$ (STO) and LaAlO$_{3}$ (LAO) have attracted much attention
\cite{Pentcheva2006,Nanda2011,Brinkman2007,Dikin2011,Huijben2009,Li2011},
particularly the prediction of charge disproportionation and the emergence
of $+3$-valent Ti sites with unpaired spin \cite{Pentcheva2006}
and direct measurements of in-plane magnetization \cite{Li2011}.
The conducting electrons at the LAO/STO interface are believed to
be induced by polar LAO's strong internal electric fields, and to
reside on the Ti sites on the STO side of the interface, partially
filling the lowest-lying Ti $3d$ bands \cite{Nakagawa2006a,Salluzzo2009,Schlom2011}.
Questions remain, however, over the role of the growth process, in
particular whether oxygen vacancy formation or cation intermixing
are in fact responsible for the observed $n$-type conduction \cite{Siemons2007,Herranz2007,Willmott2007,Kalabukhov2011a}.

Other than growing a polar overlayer, a 2D system of electrons in
STO can be made by chemical doping with Nb, La, or oxygen vacancies
\cite{Kozuka2009a,Son2010,Jalan2010,Santander-Syro2011}, or purely
electrostatic charging in an electric double layer transistor (EDLT)
\cite{Ueno2008,Lee2011b}. If electronic reconstruction in response
to overlayer polarity is an accurate description for LAO/STO, then
that system can be closely modeled by field effect-induced electrons
in undoped STO, where confounding questions over growth conditions
do not arise, and the applied electric field can be widely tuned.

In this Letter, we expand on the body of evidence for Ti$^{3+}$ magnetism
in STO that conducts in two dimensions. We demonstrate a gate-controlled
Kondo effect in the 2D electron system in undoped STO formed beneath
the bare surface by the electric field from an ionic gel electrolyte,
and interpret this system as an admixture of magnetic Ti$^{3+}$ ions
(unpaired and localized electrons) and delocalized electrons partially
filling the Ti $3d$ conduction band, as predicted theoretically \cite{Popovic2008,Nanda2011}.
The Kondo effect is an archetype for the emergent magnetic interactions
amongst localized and delocalized electrons in conducting alloys \cite{Kondo1964,Gruner1974},
and the ability to produce and tune the effect by purely electrostatic
means in any conducting system is of interest in its own right \cite{Goldhaber-Gordon1998a,Chen2011}.
The observed appearance of the Kondo effect in STO as a function of
applied electric field points to the emergence of magnetic interactions
between electrons in STO due to electron-electron correlations rather
than the presence of dopants.

\begin{figure}
\includegraphics{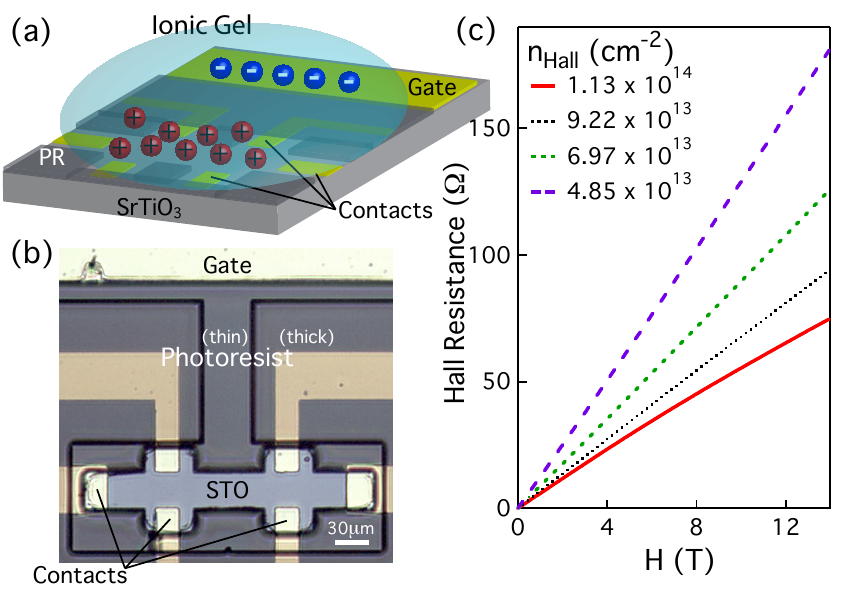}

\caption{\label{fig:1}(a) Schematic diagram of the EDLT operation. PR = photoresist.
(b) Optical micrograph of a device identical to those measured. The
photoresist (dark regions) is partially transparent, and contact leads
can be seen through it. The {}``thin'' region of photoresist has
thickness of $1\ \mu$m, while the {}``thick'' region has $2\ \mu$m.
(c) Hall resistance of device A at $T=5\mbox{ K}$, to measure the
accumulated electron density on the STO surface channel. $V_{g}=+3.5\mbox{ V}$
for the highest density, and subsequent lower densities were set by
allowing the electrolyte to partially lose polarization at $T\approx200\mbox{ K}$.}
\end{figure}

We report measurements from two STO Hall bar devices (A and B), gated
using an ionic gel electrolyte in an EDLT configuration. Behaviors
similar to those we show have been observed in 6 devices. A schematic
showing the operation of the devices is shown in Fig. \ref{fig:1}(a),
and a photograph of a device identical to those we measured but without
the electrolyte is shown in Fig. \ref{fig:1}(b). Undoped STO (100)
crystals (MTI Corp.) were treated with buffered hydrofluoric acid
to obtain a TiO$_{2}$-terminated surface \cite{Kawasaki1994}, and
the crystal for device B was then annealed at $1000\ {}^{\circ}\mbox{C}$
in a tube furnace with 50 sccm of flowing oxygen gas. The Hall bar
geometry, $30\mbox{ \ensuremath{\mu}m}$ wide and $100\mbox{ \ensuremath{\mu}m}$
long between the voltage leads, was defined via a window through a
1 $\mu$m-thick film of hard-baked photoresist that exposes the channel
and the gate to the electrolyte while keeping the rest of the STO
separated from the ions and hence still insulating. Prior to the lithographic
definition of the Hall bar, contacts were created by Ar$^{+}$ ion
milling to a dose of $2\mbox{ C/cm}^{2}$ with $300\mbox{ V}$ acceleration
\cite{Reagor2005} then depositing Al/Ti/Au electrodes with thickness
of 40/5/100 nm. The ionic gel electrolyte was formed by gelation of
a triblock copolymer poly(styrene-\emph{block}-methylmethacrylate-\emph{block}-styrene)
(PS-PMMA-PS) in an ionic liquid 1-ethyl-3-methylimidazolium bis(trifluoromethanesulfonyl)amide
(EMI-TFSA, formerly referred to as EMI-TFSI) \cite{Cho2008,Lee2009}.
A drop of gel was formed on another substrate, then manually pasted
over the device, covering both the STO channel and the $200\ \mu\mbox{m}\times400\ \mu\mbox{m}$
coplanar metal gate.

Magnetotransport characteristics of device A were measured in a Physical
Property Measurement System (Quantum Design) at temperatures down
to $T=4.5\mbox{ K}$ and magnetic fields up to $H=14\mbox{ T}$. The
sample was insulating at the start and the end of the experiment,
indicating that the conduction was not due to doping by electrochemical
reactions. At room temperature, the gate voltage was ramped up to
$V_{g}=+3.5\mbox{ V}$, which polarized the electrolyte, pushing cations
toward the channel. The electric field of the ions cause the accumulation
of electrons that form our 2D system in STO. Then the sample was cooled
to $T=5\mbox{ K}$, during which the leakage current through the gate
dropped below the measurement limit of $100\mbox{ pA}$ for $T<200\mbox{ K}$,
signaling the freezing of EMI-TFSA. Once at $T=5\mbox{ K}$, $V_{g}$
was nulled and magnetotransport measurements were taken. To apply
a weaker electric field and set the electron density lower, the device
was warmed to $T\approx200\mbox{ K}$, and the electrolyte was allowed
to partially lose its polarization, decreasing the accumulated cation
concentration at the channel and correspondingly the electron density
in the STO.

We measured the longitudinal resistance $R$ of the device as a function
of temperature and applied magnetic field, using standard lock-in
techniques at quasi-DC frequencies $<100\mbox{ Hz}$ with a current
bias $<5\mbox{ nA}$ and no additional source-drain bias. Figure \ref{fig:1}(c)
shows Hall resistance measurements at $T=5\mbox{ K}$ that show that
the electron density inferred from Hall effect decreases for each
successive cooldown, from $n_{{\rm Hall}}=1.13\times10^{14}\mbox{ cm}^{-2}$
in the first cooldown, to $n_{{\rm Hall}}=4.85\times10^{13}\mbox{ cm}^{-2}$
for the last. We have measured densities as high as $7\times10^{14}\mbox{ cm}^{-2}$
in some other samples, but the devices described here with $n_{{\rm Hall}}\sim10^{13}\mbox{ to }10^{14}\mbox{ cm}^{-2}$
showed lower disorder and have the observed density and other transport
features most similar to other high-mobility 2D systems in STO.

The 2D nature of our samples is evident from three magnetoresistance
features which have not been reported in previous EDLT studies of
STO: the dependence of the Hall resistance as a function of $H_{\perp}=H\cos\theta$,
weak anti-localization, and Shubnikov-de Haas oscillations (See Supplementary
Information). Weak anti-localization with similarly strong spin-orbit
coupling strengths has been reported in the LAO/STO interface and
attributed to the Rashba-type coupling due to inversion asymmetry
of the interface \cite{Caviglia2010a}. Shubnikov-de Haas oscillations
were seen in device B at $V_{g}=+2.8\mbox{ V}$, which gave $n_{{\rm Hall}}=2.6\times10^{13}\mbox{ cm}^{-2}$.
The electron density inferred from the oscillation period was much
lower at $3\times10^{12}\mbox{ cm}^{-2}$ if only a twofold spin degeneracy
is assumed (See Supplementary Information). Reported quantum oscillations
of the magnetoresistance in LAO/STO and $\delta$-doped STO have yielded
similarly low values of the inferred electron density \cite{Caviglia2010,BenShalom2010a,Kozuka2009a,Jalan2010}.

\begin{figure}
\includegraphics{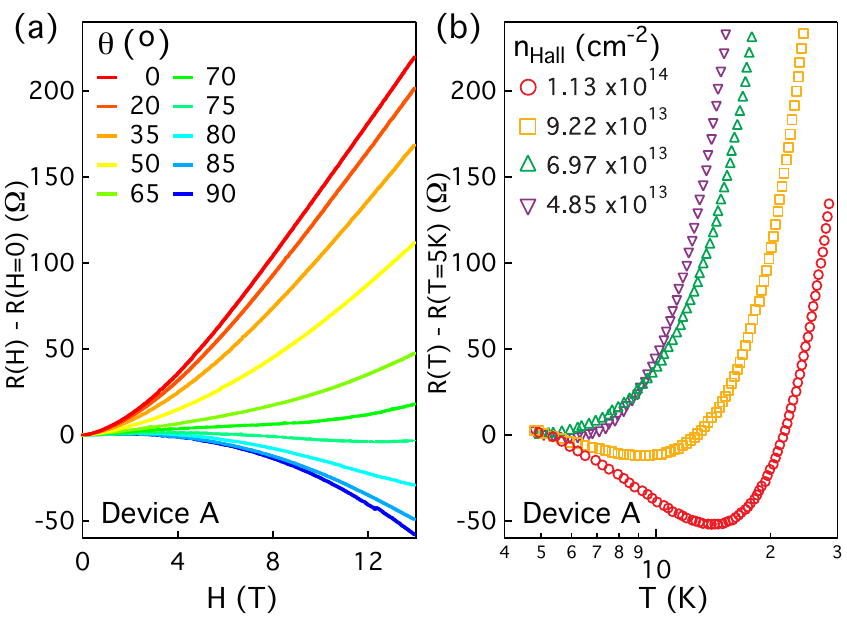}

\caption{\label{fig:2}(a) Device A magnetoresistance for various directions
of $\vec{H}$, at the highest electron density where $n_{{\rm Hall}}=1.13\times10^{14}\mbox{ cm}^{-2}$.
Angle $\theta$ is measured from the normal to sample plane, and $\theta=90^{\circ}$
is $\vec{H}\parallel\vec{j}$. $T=5\mbox{ K}$. (b) Device A $R(T)-R(T=5K)$
for each measured density in $H=0$, showing the observed $R(T)$
upturn strengthen for higher density.}
\end{figure}

The magnetoresistance of the sample, $R(H)-R(H=0)$, measured at the
highest electron density, is plotted in Fig. \ref{fig:2}(a) for various
directions of $\vec{H}$. When $\vec{H}$ is normal to the sample
plane ($\theta=0$), the magnetoresistance is positive, and as the
$\vec{H}$ direction is tilted away from normal and into the sample
plane, the magnetoresistance crosses over from positive to negative
at $\theta\approx75^{\circ}$. A similar crossover from positive out-of-plane
magnetoresistance to negative in-plane magnetoresistance has been
reported in LAO/STO \cite{Shalom2009}.

We pay particular attention to the sample's zero-field resistance
as a function of temperature, $R(T)$. Fig. \ref{fig:2}(b) shows
the resistance relative to its value at $T=5\mbox{ K}$, $R(T)-R(T=5\mbox{K})$,
at each measured electron density. At the highest density, a minimum
is seen at $T=14\mbox{ K}$ and the resistance turns upward, and this
upturn is substantially weakened as the density is lowered. The appearance
of a resistance minimum and low-temperature upturn, unexpected for
a metallic system, at higher electron density suggests that the electric
field-induced electrons, not added impurities, are themselves responsible
for the scattering, and electron-electron correlations strongly influence
the transport properties. A disorder-induced metal-insulator transition,
by contrast, ought to show a stronger upturn at lower density, not
the opposite trend seen here. The precise threshold density for the
emergence of an $R(T)$ minimum differs amongst samples, but the overall
trend is the same. Reducing the induced 2D electron density reduces
the low-temperature upturn in resistance.

To further investigate these anomalies we measured device B at lower
temperatures down to $1.4\mbox{ K}$ and higher magnetic fields up
to $31\mbox{ T}$. We set $V_{g}=+3.5\mbox{ V}$ for the first cooldown
and measured $n_{{\rm Hall}}=5.3\times10^{13}\mbox{ cm}^{-2}$ at
$T=1.4\mbox{ K}$. To set lower densities for subsequent cooldowns,
we set lower $V_{g}=+3.2,\ 2.8\mbox{, and }2.2\mbox{ V}$ at $T\approx200\mbox{ K}$
then waited for $\sim15$ minutes for the electrolyte to equilibrate,
rather than nulling $V_{g}$ and quickly obtaining a partial loss
of polarization as described above for device A.

\begin{figure}
\includegraphics{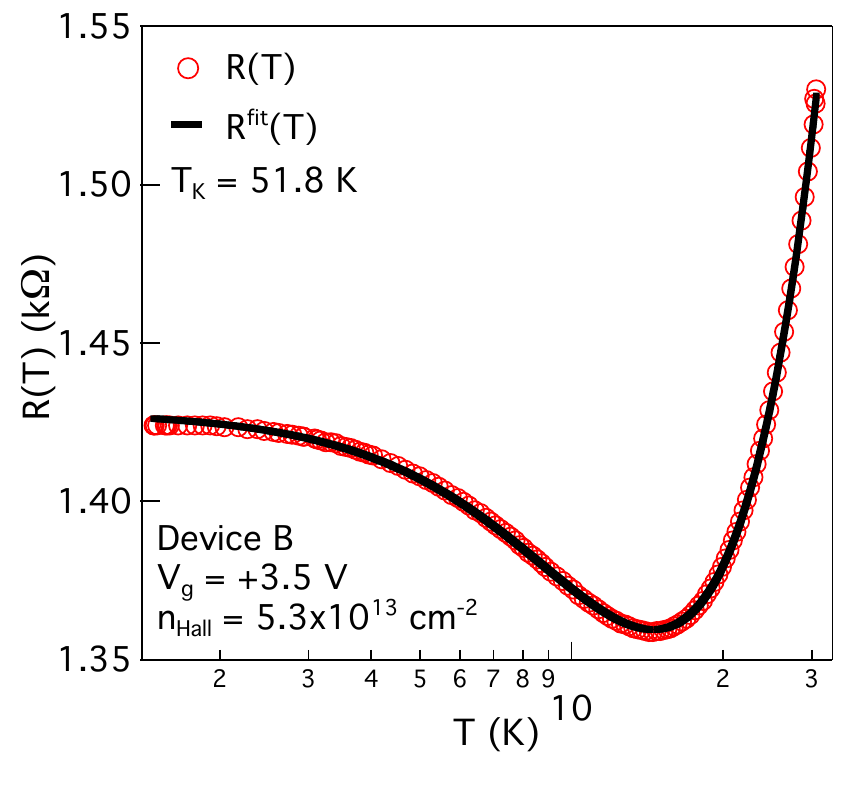}

\caption{\label{fig:3}Device B longitudinal resistance in $H=0$ as a function
of temperature at $V_{g}=+3.5\mbox{ V}$. Hall effect yielded $n_{{\rm Hall}}=5.3\times10^{13}\mbox{ cm}^{-2}$
at $T=1.4\mbox{ K}$. Solid curve: a fit using Eqns. \ref{eq:Kondo}
and \ref{eq:DGG}. $R_{0}=607\mbox{ \ensuremath{\Omega}}$, $q=0.437\ \Omega/\mbox{K}^{2}$,
$p=1.2\times10^{-8}\ \mbox{\ensuremath{\Omega}}/\mbox{K}^{5}$, $R_{K}(0\mbox{ K})=819\mbox{ \ensuremath{\Omega}}$,
and $T_{K}=51.8\mbox{ K}$.}
\end{figure}

Figure \ref{fig:3} shows the device B zero-field $R(T)$ at $V_{g}=+3.5\mbox{ V}$,
corresponding to the highest electron density measured in this sample.
A minimum of $R(T)$ is seen at $T=14.5\mbox{ K}$, and then the upturn
saturates at the lowest temperatures, such that $d^{2}R/dT^{2}<0$
for $T<7\mbox{ K}$. This saturating resistance upturn is suppressed
in subsequent cooldowns at lower gate voltages, and thus the same
overall trend in the behavior of $R(T)$ with respect to electron
density is observed in both devices. Finally in the last cooldown
and lowest density, a non-saturating upturn and localization-like
behavior is observed (See Supplementary Information).

The appearance of a saturating resistance upturn at low temperature
is characteristic of the Kondo effect, where the temperature dependence
of the contribution from magnetic impurities to the electrical resistivity
of a metal is a universal function in units of a single temperature
scale, the Kondo temperature $T_{K}$. This universal function $R_{K}\left(T/T_{K}\right)$
behaves logarithmically at high temperature $T\gg T_{K}$, and saturates
at low temperature, so that $R_{K}\left(T/T_{K}\right)\approx R_{K}(0\mbox{ K})\left(1-6.088\left(T/T_{K}\right)^{2}\right)$
for $T\ll T_{K}$ if we define $T_{K}$ as the temperature at which
the Kondo resistivity is half relative to its zero-temperature value
\cite{Kondo1964,Costi1994}. Across the whole measured temperature
range from $1.4\mbox{ K}$ to $30\mbox{ K}$ which includes temperatures
above and below the observed $R(T)$ minimum, the resistance can be
described well by a simple Kondo model 
\begin{equation}
R^{{\rm fit}}\left(T\right)=R_{0}+qT^{2}+pT^{5}+R_{K}\left(T/T_{K}\right)\label{eq:Kondo}
\end{equation}
 where $R_{0}$ is the residual resistance due to sample disorder,
and the $T^{2}$ and $T^{5}$ terms represent the functional temperature
dependences of the electron-electron and the electron-phonon interactions,
respectively. For the numerical fitting of this model to the data,
we used an empirical form for the universal resistivity function,
\begin{equation}
R_{K}\left(T/T_{K}\right)=R_{K}\left(0\mbox{ K}\right)\left(\frac{T_{K}'^{2}}{T^{2}+T_{K}'^{2}}\right)^{s}\label{eq:DGG}
\end{equation}
 where $T_{K}'=T_{K}/\left(2^{1/s}-1\right)^{1/2}$, and we fixed
$s=0.225$ to closely match the theoretical result obtained by the
numerical renormalization group \cite{Costi1994,Goldhaber-Gordon1998}.
A numerical fit using Eqns. \ref{eq:Kondo} and \ref{eq:DGG} to the
measured $R(T)$ curve yielded $R_{0}=607\mbox{ \ensuremath{\Omega}}$,
$q=0.437\ \Omega/\mbox{K}^{2}$, $p=1.2\times10^{-8}\ \mbox{\ensuremath{\Omega}}/\mbox{K}^{5}$,
$R_{K}(0\mbox{ K})=819\mbox{ \ensuremath{\Omega}}$, and $T_{K}=51.8\mbox{ K}$.

\begin{figure}
\includegraphics{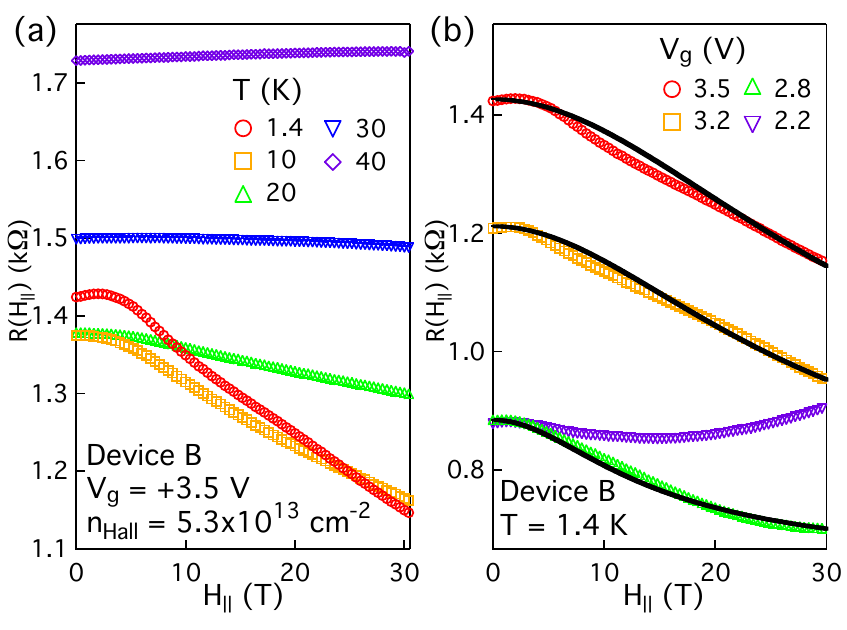}

\caption{\label{fig:4}(a) Device B in-plane magnetoresistance ($\vec{H}\perp\vec{j}$)
at $V_{g}=+3.5\mbox{ V}$ and various temperatures. The negative magnetoresistance
is gradually suppressed as the temperature is raised, and the resistance
increases with temperature in the highest $H_{\parallel}$. (b) The
same at all four applied gate voltages at $T=1.4\mbox{ K}$. Solid
curves: $R^{{\rm model}}\left(H_{\parallel}\right)$ according to
Eqn. \ref{eq:Bethe} where we chose $R_{0}=607\mbox{ \ensuremath{\Omega}}$,
and $H_{1}=20,\ 16\mbox{, and }8\mbox{ T}$ for $V_{g}=+3.5,\ 3.2\mbox{,\ and }2.8\mbox{ V}$,
respectively.}
\end{figure}

Device B, with $\vec{H}$ applied parallel to sample plane and $\vec{H}\perp\vec{j}$,
also exhibited a strong negative magnetoresistance, up to $\approx-20\%$
at $H_{\parallel}=31\mbox{ T}$, and its $R\left(H_{\parallel}\right)$
is plotted in Fig. \ref{fig:4}. The temperature-dependent magnetoresistance
at $V_{g}=+3.5\mbox{ V}$, plotted in Fig. \ref{fig:4}(a), shows
that raising the sample temperature suppresses the negative in-plane
magnetoresistance such that the effect disappears between 30 and $40\mbox{ K}$.
In the highest fields $H_{\parallel}>25\mbox{ T}$, the Kondo resistance
upturn is sufficiently suppressed that resistance depends monotonically
on temperature. Such a temperature dependence of the magnetoresistance
is expected from the splitting of the Kondo Peak in the spectral function
by an applied magnetic field, which leads to negative magnetoresistance
by suppressing the Kondo effect on resistance \cite{Felsch1973,Costi2000}.

Measurements of $R\left(H_{\parallel}\right)$ were done at each gate
voltage at the lowest temperature $T=1.4\mbox{ K}$, and the results
are plotted in Fig. \ref{fig:4}(b) along with their comparison to
a simple Kondo model 
\begin{equation}
R^{{\rm model}}\left(H_{\parallel}\right)=R_{0}+R_{K}\left(H_{\parallel}/H_{1}\right)\label{eq:Bethe}
\end{equation}
 where now $R_{K}\left(H_{\parallel}/H_{1}\right)$ is the universal
function for zero-temperature magnetoresistivity of a Kondo impurity,
related to its magnetization which may be calculated using Bethe-Ansatz
techniques, and $H_{1}$ is a magnetic scale, related to the Kondo
temperature and the $g$-factor of the impurity spin \cite{Andrei1983}.
To compare the data and model we numerically evaluated the exact form
of $R_{K}\left(H_{\parallel}/H_{1}\right)$ (See Supplementary Information
and \cite{Andrei1983}) and chose $H_{1}=20,\ 16\mbox{, and }8\mbox{ T}$
to compare with the data obtained at $V_{g}=+3.5,\ 3.2\mbox{,\ and }2.8\mbox{ V}$,
respectively. The comparison is not exact and fails for low fields,
where weak anti-localization and other effects may be relevant, but
the overall dependence and shape are broadly consistent, particularly
for the lower density at $V_{g}=+2.8\mbox{ V}$.

The negative in-plane magnetoresistance is suppressed as the electron
density is lowered, once again implying that the observed Kondo effect
originates from the electrons accumulated at the surface due to the
applied electric field of the electrolyte gate. As the Kondo effect
arises due to the interaction between localized and delocalized electrons,
we can interpret our observations by viewing the 2D electron system
as an admixture composed of localized and unpaired electrons---perhaps
polaronic in nature \cite{Nanda2011}---that act as the Kondo scattering
centers and a metal of delocalized electrons that partially fill the
Ti $3d$ conduction band.

Finally, we comment on the implications of the results described herein
for the ongoing efforts to conclusively understand the LAO/STO interface,
where an $R(T)$ minimum and negative magnetoresistance have already
been reported \cite{Brinkman2007}. By electrostatically inducing
2D electrons in STO, we have modeled the essential physics of the
polar catastrophe, and the demonstration of a gate-controlled Kondo
effect in undoped STO shows that the observations of electronic conduction
and magnetism in LAO/STO are plausibly due to electronic reconstruction
and are not necessarily a result of unintended doping during LAO growth
\cite{Huijben2009,Schlom2011}. The magnetic interactions found in
STO, added to its other attractive features including tunability of
the ground state by applied electric fields and superconductivity,
show STO-based interfaces and heterostructures to be a promising playground
in which to look for and study emergent electronic phenomena and novel
device applications \cite{Takagi2010a,Mannhart2010}.
\begin{acknowledgments}
We thank J. Jaroszynski, S. Stemmer, B. Jalan, I. R. Fisher, J. G.
Analytis, M. Chalfin, E. Eason, Y. Cui, J. J. Cha, Y. Lee, and T.
A. Costi. The development of ionic gating technique was supported
as part of the Center on Nanostructuring for Efficient Energy Conversion,
an Energy Frontier Research Center funded by the U.S. Department of
Energy, Office of Science, Office of Basic Energy Sciences under Award
Number DE-SC0001060. The measurement and study of STO were supported
by the MURI program of the Army Research Office Grant No. W911-NF-09-1-0398.
The Minnesota contribution was supported by the National Science Foundation
through the MRSEC program at the University of Minnesota, Award DMR-0819885.
 ML is partially supported by Samsung Scholarship and the Stanford
Graduate Fellowship. A portion of this work was performed at the National
High Magnetic Field Laboratory, which is supported by National Science
Foundation Cooperative Agreement No. DMR-0654118, the State of Florida,
and the U.S. Department of Energy.\end{acknowledgments}


\begin{thebibliography}{References}
\bibitem{Pentcheva2006}R. Pentcheva and W. Pickett, Phys. Rev. B
\textbf{74} 035112 (2006).

\bibitem{Nanda2011}B. Nanda and S. Satpathy, Phys. Rev B \textbf{83},
195114 (2011).

\bibitem{Brinkman2007}A. Brinkman \emph{et al}., Nat. Mater. \textbf{6},
493 (2007).

\bibitem{Dikin2011}D. A. Dikin \emph{et al}., Phy. Rev. Lett. \textbf{107},
056802 (2011).

\bibitem{Huijben2009}M. Huijben \emph{et al}., Adv. Mater. \textbf{21},
1665 (2009).

\bibitem{Li2011}L. Li \emph{et al}., arXiv:1105.0235.

\bibitem{Nakagawa2006a}N. Nakagawa, H. Y. Hwang, and D. A. Muller,
Nat. Mater. \textbf{5}, 204 (2006).

\bibitem{Salluzzo2009}M. Salluzzo \emph{et al}., Phys. Rev Lett.
\textbf{102}, 166804 (2009).

\bibitem{Schlom2011}D. G. Schlom and J. Mannhart, Nat. Mater. \textbf{10},
168 (2011).

\bibitem{Siemons2007}W. Siemons \emph{et al}., Phys. Rev Lett. \textbf{98},
196802 (2007).

\bibitem{Herranz2007}G. Herranz \emph{et al}., Phys. Rev. Lett. \textbf{98},
216803 (2007).

\bibitem{Willmott2007}P. Willmott \emph{et al}., Phys. Rev. Lett.
\textbf{99}, 155502 (2007).

\bibitem{Kalabukhov2011a}A. Kalabukhov \emph{et al}., Europhys. Lett.
\textbf{93}, 37001 (2011).

\bibitem{Kozuka2009a}Y. Kozuka \emph{et al}., Nature \textbf{462},
487 (2009).

\bibitem{Son2010}J. Son \emph{et al}., Nat. Mater. \textbf{9}, 482
(2010).

\bibitem{Jalan2010}B. Jalan \emph{et al}., Phys. Rev. B \textbf{82}
081103 (2010).

\bibitem{Santander-Syro2011}A. F. Santander-Syro \emph{et al}., Nature
\textbf{469}, 189 (2011).

\bibitem{Ueno2008}K. Ueno \emph{et al}., Nat. Mater. \textbf{7},
855 (2008).

\bibitem{Lee2011b}Y. Lee \emph{et al}., Phys. Rev. Lett. \textbf{106},
136809 (2011).

\bibitem{Popovic2008}Z. Popovic, S. Satpathy, and R. Martin, Phys.
Rev. Lett. \textbf{101}, 256801 (2008).

\bibitem{Kondo1964}J. Kondo, Prog. Theor. Phys. \textbf{32}, 37 (1964).

\bibitem{Gruner1974}G. Gruner and A. Zawadowski, Reports on Prog.
in Physics \textbf{37}, 1497 (1974).

\bibitem{Goldhaber-Gordon1998a}D. Goldhaber-Gordon \emph{et al}.,
Nature \textbf{391}, 156 (1998).

\bibitem{Chen2011}J.-H. Chen \emph{et al}., Nat. Phys. \textbf{7},
535 (2011).

\bibitem{Kawasaki1994}M. Kawasaki \emph{et al}., Science \textbf{266},
1540 (1994).

\bibitem{Reagor2005}D. W. Reagor and V. Y. Butko. Nat. Mater. \textbf{4},
593 (2005).

\bibitem{Cho2008}J. H. Cho \emph{et al}., Nat. Mater. \textbf{7},
900 (2008).

\bibitem{Lee2009}J. Lee \emph{et al}., J. Phys. Chem. C \textbf{113},
8972 (2009).

\bibitem{Caviglia2010a}A. D. Caviglia \emph{et al}., Phys. Rev. Lett.
\textbf{104}, 126803 (2010).

\bibitem{Caviglia2010}A. D. Caviglia \emph{et al}., Phys. Rev. Lett.
\textbf{105}, 236802 (2010).

\bibitem{BenShalom2010a}M. Ben Shalom \emph{et al}., Phys. Rev. Lett.
\textbf{105}, 206401 (2010).

\bibitem{Shalom2009}M. Ben Shalom \emph{et al}., Phys. Rev. B \textbf{80},
140403 (2009).

\bibitem{Costi1994}T. A. Costi, A. C. Hewson, and V. Zlatic, J. of
Phys.: Condens. Matter \textbf{6}, 2519 (1994).

\bibitem{Goldhaber-Gordon1998}D. Goldhaber-Gordon \emph{et al}.,
Phys. Rev. Lett. \textbf{81}, 5225 (1998).

\bibitem{Felsch1973}W. Felsch and K. Winzer, Solid State Commun.
\textbf{13}, 569 (1973).

\bibitem{Costi2000}T. Costi., Phys. Rev. Lett. \textbf{85}, 1504
(2000).

\bibitem{Andrei1983}N. Andrei, K. Furuya, and J. Lowenstein, Rev.
Mod. Phys. \textbf{55}, 331 (1983).

\bibitem{Takagi2010a}H. Takagi and H. Y. Hwang, Science \textbf{327},
1601 (2010).

\bibitem{Mannhart2010}J. Mannhart and D. G. Schlom, Science \textbf{327},
1607 (2010).\end{thebibliography}
\end{document}

\title{Supplementary Information for {}``Electrolyte gate-controlled Kondo
effect in SrTiO$_{3}$''}

\author{Menyoung Lee}

\affiliation{Department of Physics, Stanford University, Stanford, California
94305, USA}

\author{J. R. Williams}

\affiliation{Department of Physics, Stanford University, Stanford, California
94305, USA}

\author{Sipei Zhang}

\affiliation{Department of Chemical Engineering and Materials Science, University
of Minnesota, Minneapolis, Minnesota 55455, USA}

\author{C. Daniel Frisbie}

\affiliation{Department of Chemical Engineering and Materials Science, University
of Minnesota, Minneapolis, Minnesota 55455, USA}

\author{D. Goldhaber-Gordon}

\email{goldhaber-gordon@stanford.edu}

\affiliation{Department of Physics, Stanford University, Stanford, California
94305, USA}

\date{August 4, 2011}

\maketitle

\section{Two-dimensionality}

Several magnetotransport features of our devices show the two-dimensionality
of the STO electron system that is measured. These transport properties
that demonstrate two-dimensionality have not been shown in previous
reports of electrolyte-gated STO.

\subsection{Hall resistance non-linearity and multiband transport in 2D}

\begin{center}
\begin{figure}
\begin{centering}
\includegraphics[width=1\columnwidth]{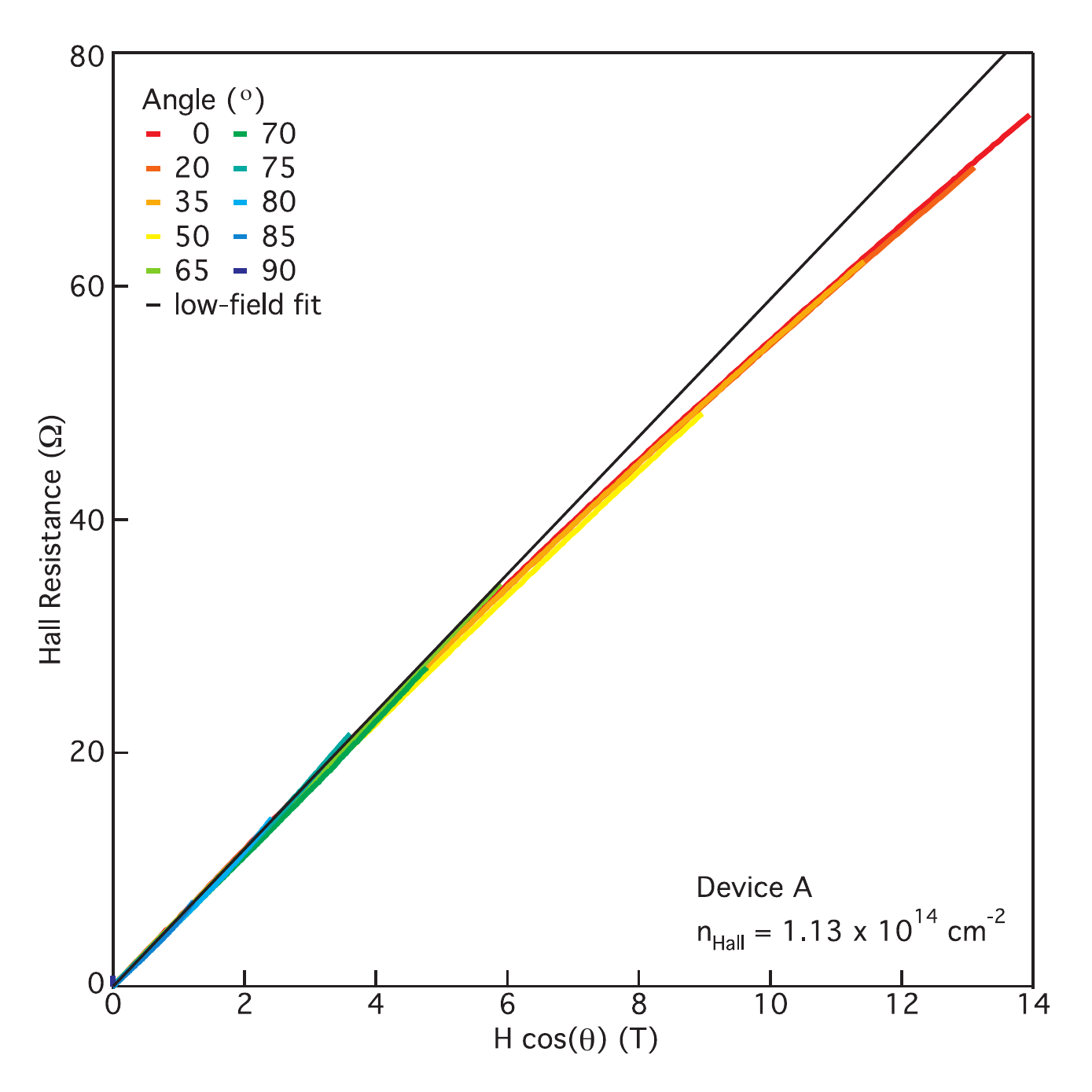}
\par\end{centering}

\caption{\label{fig:1}Device A Hall resistance plotted against $H_{\perp}=H\cos\theta$.}
\end{figure}

\par\end{center}

Figure \ref{fig:1} shows the Hall resistance of device A, at the
highest density $n$ at $T=5\mbox{ K}$ as a function of the applied
magnetic field's component perpendicular to sample plane, $H_{\perp}=H\cos\theta$.
There is a small nonlinearity and a departure from the linear fit
to the low-field trend. This departure occur at a fixed value of $H_{\perp}$,
not of $H$, and the entire nonlinear Hall resistance curve is a function
of $H_{\perp}$ rather than of $H$. This $H_{\perp}$ dependence
of a weak nonlinearity in the Hall resistance suggests multiband transport
in two dimensions.

\subsection{Weak anti-localization}

\begin{center}
\begin{figure}
\begin{centering}
\includegraphics[width=1\columnwidth]{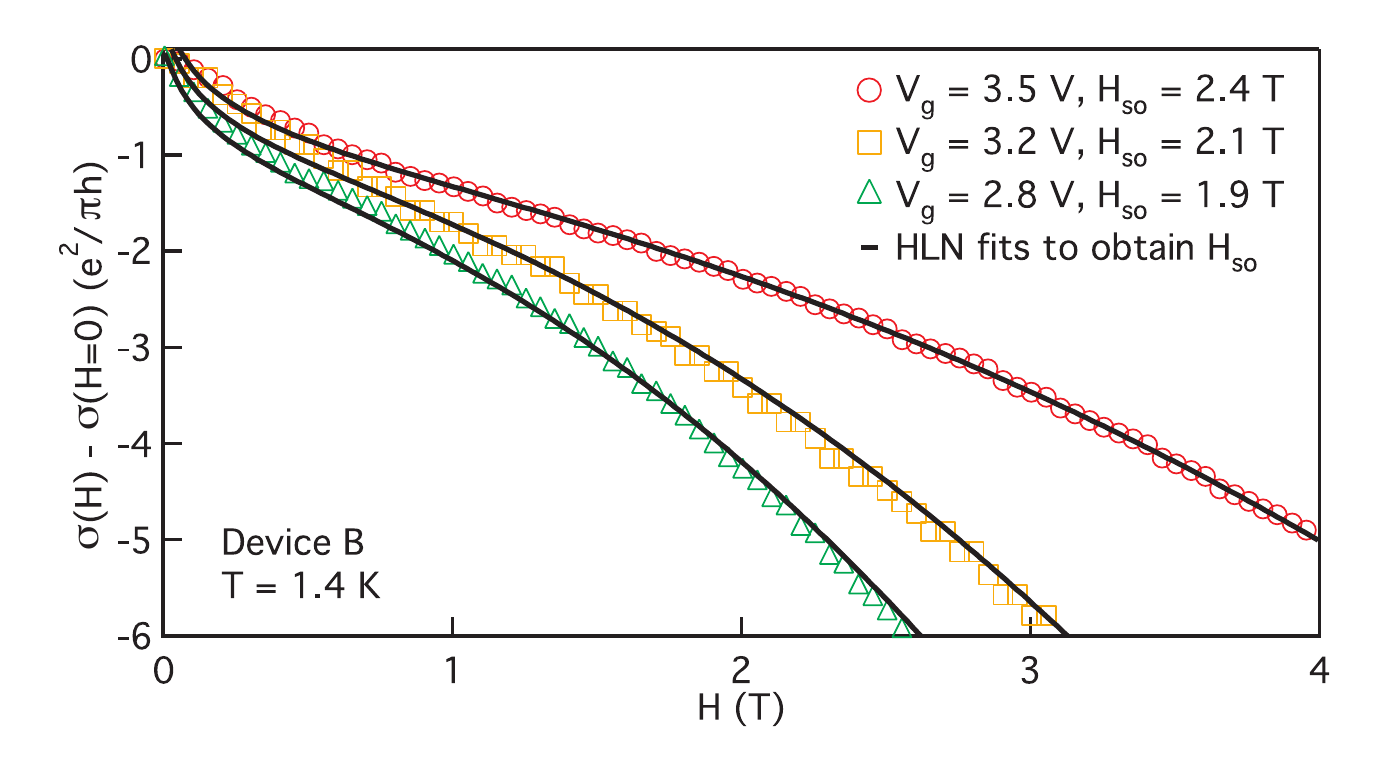}
\par\end{centering}

\caption{\label{fig:2}Device B sheet magnetoconductance at various values
of $V_{g}$. $H$ is applied normal to sample plane, and $T=1.4\mbox{ K}$.
Solid lines are a fit to the theory in \cite{Hkami1980} and \cite{Hansen1993},
which assumed that the spin-orbit coupling is isotropic and described
by a single spin-orbit field scale $H_{so}$, and that $V_{g}$ does
not change the electron scattering time $\tau_{e}$ or dephasing time
$\tau_{\phi}$ .}
\end{figure}

\par\end{center}

Figure \ref{fig:2} shows device $B$ sheet conductance $\sigma=LW^{-1}R^{-1}$,
where $L$ and $W$ are the length ($100\ \mu$m, the longitudinal
distance between the midpoints of voltage-probing contacts), and the
width ($30\ \mu$m) of the Hall bar channel, and $R$ is the longitudinal
resistance, for the first three cooldowns with $V_{g}=+3.5,\ +3.2,\mbox{ and +}2.8\mbox{ V}$,
as a function of applied magnetic field perpendicular to the sample
plane. The conductance peak at zero field has a magnitude on the order
of $e^{2}/h$, suggesting weak anti-localization as the mechanism
that generates the conductance peak.

The calculation due to Hikami, Larkin, and Nagaoka predicts a weak
(anti-)localization contribution to the low-field magnetoconductance
given by \cite{Hkami1980,Hansen1993}: 

\begin{align*}
\Delta\sigma\left(H\right) & =\frac{e^{2}}{\pi h}\left(\psi\left(\frac{1}{2}+\frac{H_{1}}{H}\right)-\psi\left(\frac{1}{2}+\frac{H_{2}}{H}\right)\right.\\
 & \qquad\left.+\frac{1}{2}\psi\left(\frac{1}{2}+\frac{H_{3}}{H}\right)-\frac{1}{2}\psi\left(\frac{1}{2}+\frac{H_{4}}{H}\right)\right)
\end{align*}
 where $\psi(x)$ is the digamma function, and the field scale fit
parameters can be written in terms of characteristic fields as $H_{1}=H_{\phi}+H_{so}^{x}+H_{so}^{y}+H_{so}^{z}$,
$H_{2}=H_{e}$, $H_{3}=H_{\phi}+2H_{so}^{x}+2H_{so}^{y}$, and $H_{4}=H_{\phi}$,
and these characteristic fields can in turn be written as $H_{e}=\frac{h}{8\pi eD\tau_{e}}$,
$H_{\phi}=\frac{h}{8\pi eD\tau_{\phi}}$, and $H_{so}^{x,y,z}=\frac{h}{8\pi eD\tau_{so}^{x,y,z}}$
in terms of the electron scattering and dephasing times $\tau_{e,\phi}$
as well as the anisotropic spin-orbit scattering times $\tau_{so}^{x,y,z}$.
For the numerical fit we have added the above $\Delta\sigma$ function
to a quadratic background $-aH^{2}$, and constrained the parameters
by assuming an isotropic spin-orbit coupling so that $H_{so}=3H_{so}^{x}=3H_{so}^{y}=\frac{3}{2}H_{so}^{z}$
and hence $H_{1}=H_{3}=H_{so}+H_{\phi}$. Furthermore we surmise that
the disorder landscape is not altered greatly in the range of electron
densities explored, so that $\tau_{e}$, $\tau_{\phi}$, and hence
$H_{2,4}$, may be held to be independent of electron density. The
solid lines in Fig. \ref{fig:2} show the resulting numerical fit,
with $H_{e}=31\mbox{ T}$, $H_{\phi}=6\mbox{ mT}$, and the inferred
spin-orbit coupling field strengths of $H_{so}=2.4,\ 2.2,\mbox{ and }1.9\mbox{ T}$
for $V_{g}=+3.5,\ +3.2,\mbox{ and +}2.8\mbox{ V}$, respectively.

Weak anti-localization with similarly strong spin-orbit coupling strengths
in the few Tesla range has been reported in the LAO/STO interface
and attributed to the Rashba-type coupling due to inversion asymmetry
of the interface \cite{Caviglia2010a}. A value of $H_{e}$ in the
few tens of Tesla range is consistent with the appearance of Shubnikov-de
Haas oscillations at a similar magnitude of applied field, which we
discuss next.

\subsection{Shubnikov-de Haas oscillations}

\begin{center}
\begin{figure}
\begin{centering}
\includegraphics[width=1\columnwidth]{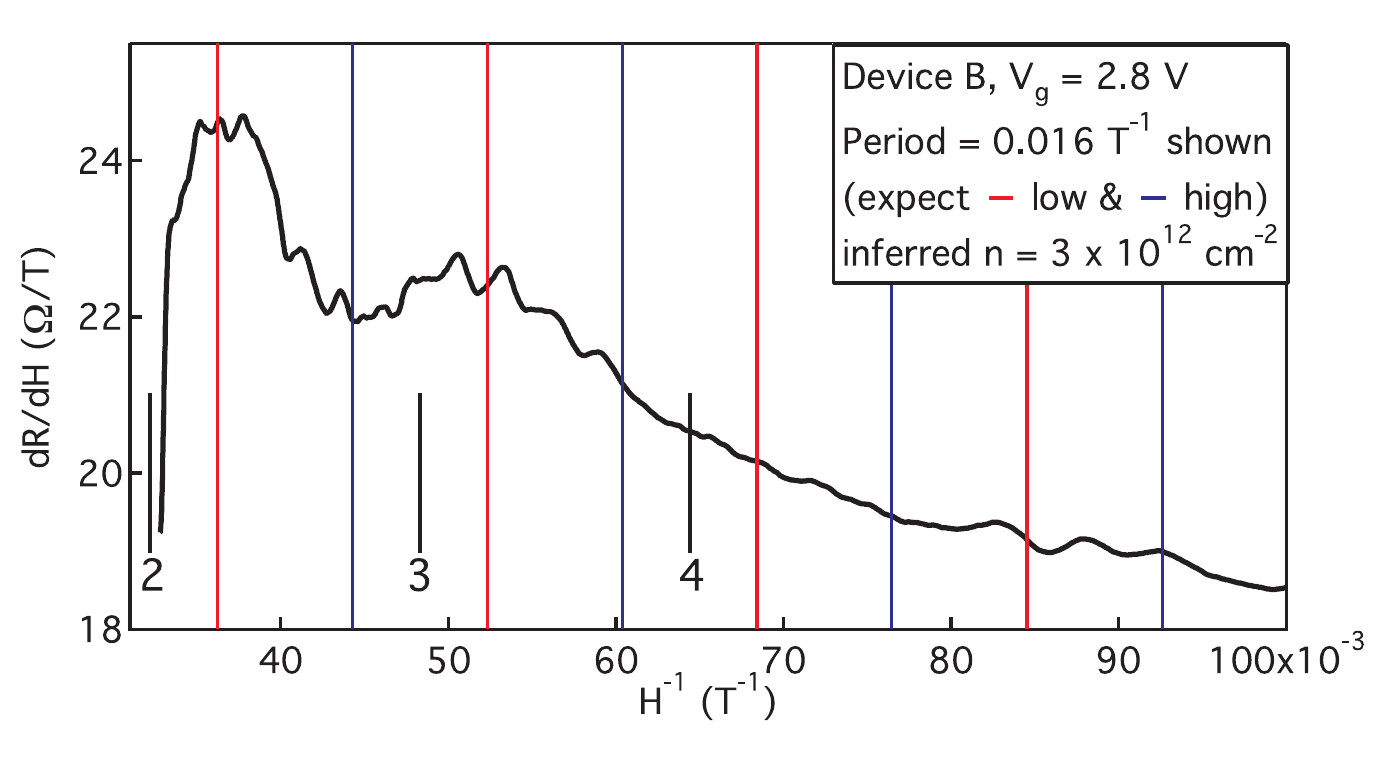}
\par\end{centering}

\caption{\label{fig:3}Shubnikov-de Haas oscillations in device $B$ magnetoresistance,
at $V_{g}=+2.8\mbox{ V}$ and $T=1.4\mbox{ K}$, with the magnetic
field applied normal to sample plane. Plotted is $dR/dH$ vs $H^{-1}$.
Red and blue vertical lines are where $dR/dH$ minima and maxima,
respectively, are expected for a simple normal 2D metal with a sheet
carrier density of $n=3\times10^{12}\mbox{ cm}^{-2}$ and corresponding
Shubnikov-de Haas oscillation period of $0.016\mbox{ T}^{-1}$. The
observed oscillations have an inverted amplitude relative to the expectation.
The short black vertical lines with numbers indicate the number of
Landau levels filled and the fields at which $\nu=2$, 3, and 4.}
\end{figure}

\par\end{center}

We have observed Shubnikov-de Haas oscillations in device $B$ magnetoresistance,
at $V_{g}=+2.8\mbox{ V}$, $T=1.4\mbox{ K}$, with the magnetic field
applied normal to sample plane. Figure \ref{fig:3} shows the plot
of the derivative of the resistance with respect to magnetic field,
$dR/dH$, as a function of the inverse of the strength of applied
magnetic field, $H^{-1}$. Only a few maxima and minima can be resolved,
but the features do seem to appear regularly in inverse magnetic field,
as expected for Shubnikov-de Haas oscillations. The amplitude of the
oscillations, however, appear to be inverted from the expected sign.
The oscillations in inverse magnetic field show a period of $0.016\mbox{ T}^{-1}$,
which corresponds to an inferred 2D electron density of $3\times10^{12}\mbox{ cm}^{-2}$
if we account for a spin degeneracy of 2 and no other sources of degeneracy.
This value is much lower than the density inferred from the Hall effect,
$n_{{\rm Hall}}=2.6\times10^{13}\mbox{ cm}^{-2}$, as also that observed
in other STO systems \cite{Jalan2010,Caviglia2010,BenShalom2010a}.

\section{Kondo effect vs localization}

\begin{center}
\begin{figure}
\begin{centering}
\includegraphics[width=1\columnwidth]{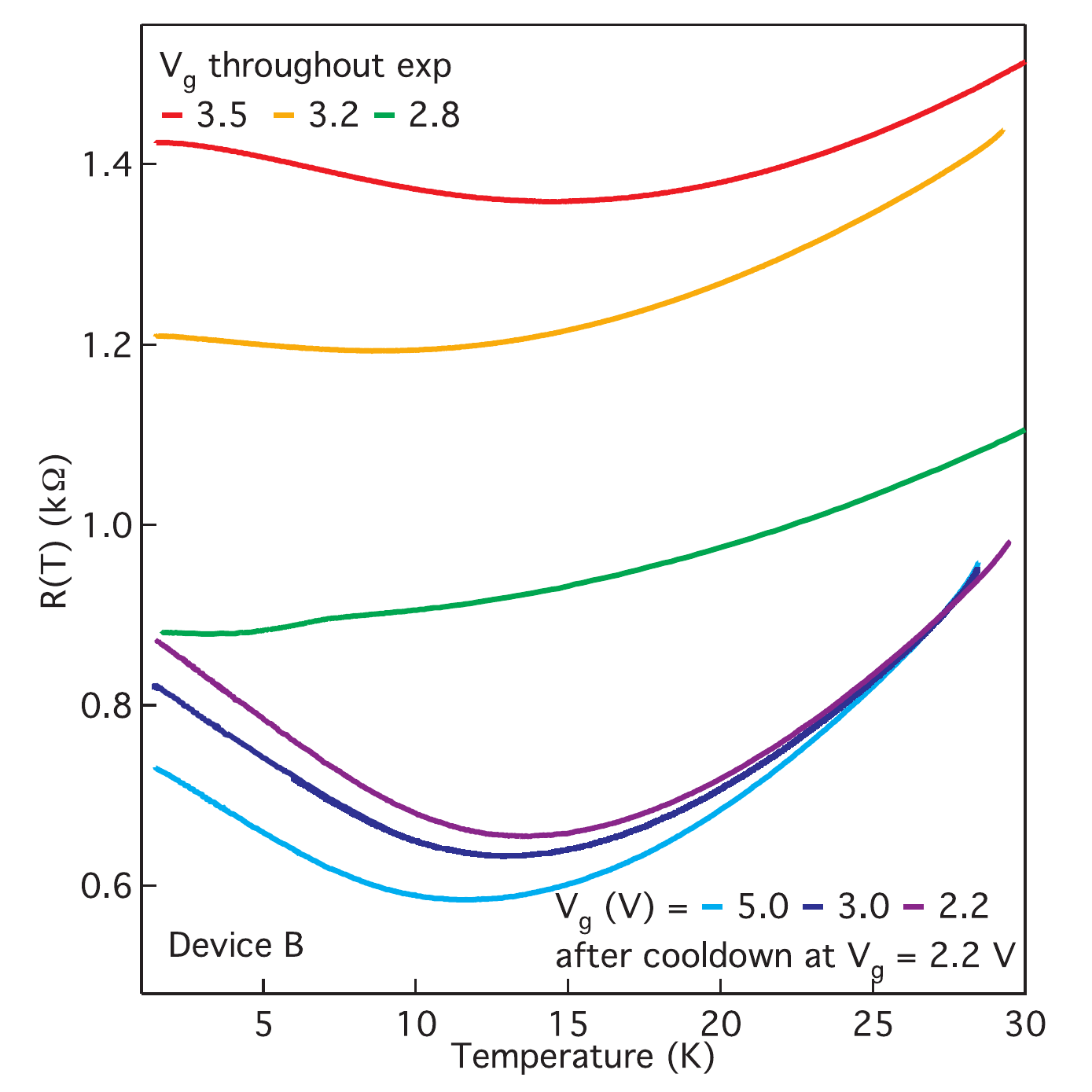}
\par\end{centering}

\caption{\label{fig:4}Device B $R(T)$ at various gate voltages $V_{g}$.
For $V_{g}=+3.5,\ +3.2,\mbox{ and +}2.8\mbox{ V}$, the gate voltage
was set at high temperature, and kept at that value during the entire
cooldown. For the final, fourth cooldown with $V_{g}$ set at $+2.2\mbox{ V}$
at high temperature, the gate voltage was further manipulated at low
temperatures with the electrolytes frozen in place.}
\end{figure}

\par\end{center}

By changing the gate voltage $V_{g}$ of the ionic gel electrolyte
gate in device B, we can tune the density of the 2D conducting system
in STO, from the highest density cooldown at $V_{g}=+3.5\mbox{ V}$
where we see the Kondo effect and its saturating resistance upturn
clearly to the lowest density cooldown at $V_{g}=+2.2\mbox{ V}$ where
a metal-insulator transition occurs and electrons localize as the
temperature is lowered. Plots of $R(T)$ at the various values of
$V_{g}$ are shown in Figure \ref{fig:4}.

The two distinct mechanisms, Kondo and localization, that give rise
to a minimum in $R(T)$ can be distingished by the fact that the resistance
upturn due to Kondo effect saturates, and $d^{2}R/dT^{2}<0$ at the
lowest temperatures (for $T<7\mbox{ K}$ at $V_{g}=+3.5\mbox{ V}$
in our device). A localization-induced resistance upturn, in contrast,
shows no saturation and $d^{2}R/dT^{2}>0$ for all $T$. This non-saturation
of upturn holds for $R(T)$ across the whole measured temperature
range for the cooldown at $V_{g}=+2.2\mbox{ V}$.

For the lowest-density cooldown at $V_{g}=+2.2\mbox{ V}$, we found
in addition that varying the gate voltage at low temperatures where
the electrolyte is frozen in place had an additional effect on the
sample resistance. In this case the gate voltage on our coplanar electrode
acts as a dielectric gate simlar to a back gate at the bottom of the
STO substrate. The trend of the gate's effect at low temperature was
conventional, i.e., increasing the low-temperature gate voltage decreased
the resistance as well as the temperature at which the localization
effect becomes apparent. Increasing $V_{g}$ once at low temperature
from $+2.2$ to $+3.0$ and $+5.0\mbox{ V}$ decreased the low-temperature
resistance of the sample, and shifted the minimum of the $R(T)$ minimum
to lower temperature, indicating the approach of a metal-insulator
transition from the localized, insulating side.

\section{Kondo magnetoresistance}

\begin{center}
\begin{figure}
\begin{centering}
\includegraphics[width=1\columnwidth]{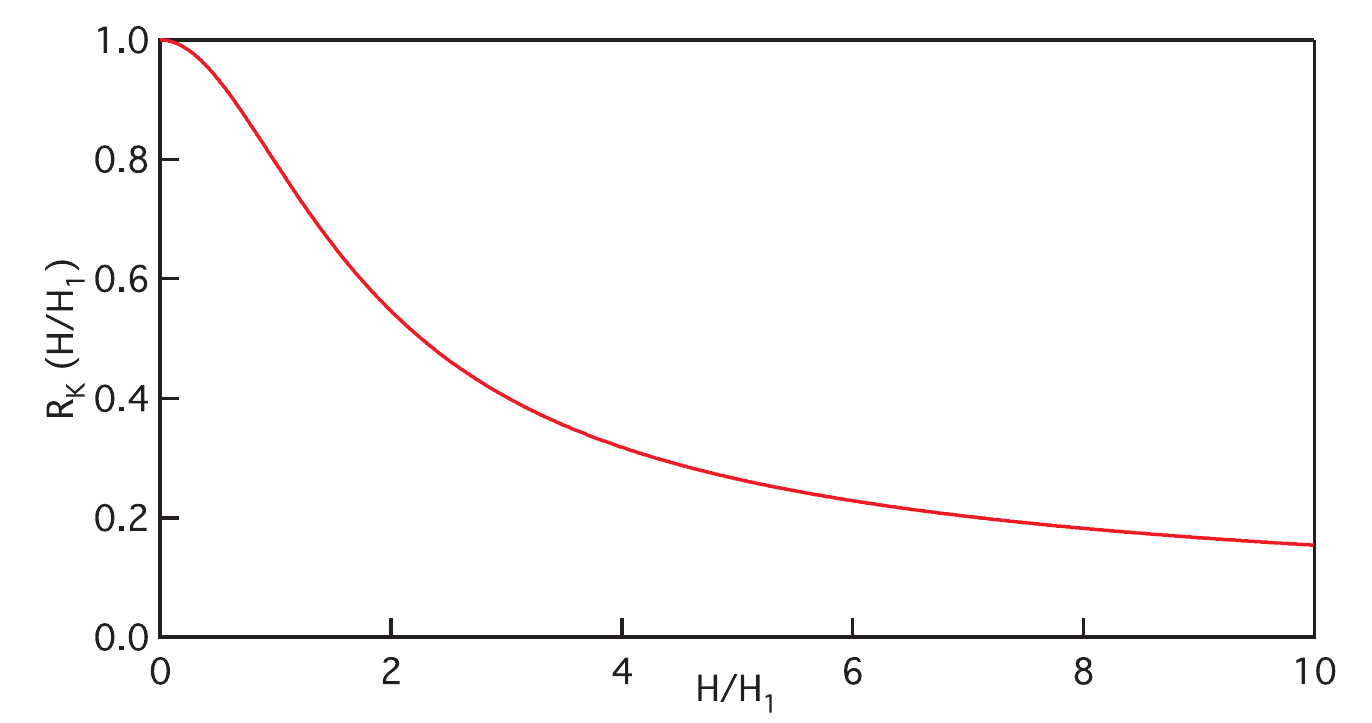}
\par\end{centering}

\caption{\label{fig:5}Universal zero-temperature magnetoresistance due to
a Kondo impurity: plot of the exact solution obtained by the Bethe-Ansatz
method in \cite{Andrei1983}.}
\end{figure}

\par\end{center}

To compare our sample in-plane magnetoresistance with the available
theory, we calculated the zero-temperature Kondo magnetoresistance
obtained by Bethe-Ansatz methods in \cite{Andrei1983}.

At zero temperature, a Kondo impurity's magnetization is given by

\[
M\left(H/H_{1}\right)=\begin{cases}
\frac{1}{\sqrt{2\pi}}\sum_{k=0}^{\infty}\left(-\frac{1}{2}\right)^{k}\left(k!\right)^{-1}\left(k+\frac{1}{2}\right)^{k-\frac{1}{2}}\\
\qquad e^{-\left(k+\frac{1}{2}\right)}\left(\frac{H}{H_{1}}\right)^{2k+1},\ H\leq\sqrt{2}H_{1}\\
1-\pi^{-3/2}\int\frac{dt}{t}\sin\left(\pi t\right)e^{-t\ln\left(t/2e\right)}\\
\qquad\left(\frac{H_{1}}{H}\right)\Gamma\left(t+\frac{1}{2}\right),\ \sqrt{2}H_{1}\leq H
\end{cases}
\]
 where $H_{1}$ is a magnetic field scale, related to both the Kondo
temperature and the g-factor of the impurity spin. Knowing the magnetization
of the impurity in turn yields the zero-temperature Kondo magnetoresistance
by the relation 
\[
R_{K}\left(H/H_{1}\right)=R_{K}\left(H=0\right)\cos^{2}\left(\frac{\pi}{2}M\left(H/H_{1}\right)\right).
\]

We plot this standard curve in Figure \ref{fig:5}. Figure 4(b) of
the main text shows the comparison of measured device B $R(H_{\parallel})$
against this standard curve, scaled and offset.